\documentclass[final]{ws-p8-50x6-00}

\begin{document}
\def\lsi{\raise0.3ex\hbox{$<$\kern-0.75em\raise-1.1ex\hbox{$\sim$}}}
\def\gsi{\raise0.3ex\hbox{$>$\kern-0.75em\raise-1.1ex\hbox{$\sim$}}}
\newcommand{\lsim}{\mathop{\lsi}}
\newcommand{\gsim}{\mathop{\gsi}}

\title{Preheating and phase transitions\\ in gauge theories
\footnote{
T\lowercase{alk presented at} SEWM2000, 
M\lowercase{arseille}, 
F\lowercase{rance}, 14--17 
J\lowercase{une}, 2000.}
}

\author{A.~Rajantie}

\address{Centre for Theoretical Physics, University of Sussex,
Brighton BN1 9QH, UK
\footnote{Present address:
DAMTP, CMS, Wilberforce Road, Cambridge CB3 0WA, UK
}
}

\maketitle

\abstracts{
It has recently been suggested that the baryon washout problem of the
standard electroweak baryogenesis scenario
could be avoided if inflation ends with a period of parametric resonance
at a low enough energy density.
I present results of numerical simulations in which 
this process was studied in the Abelian Higgs model.
Our results show that because of the masslessness of the gauge
field, the parametric resonance takes place naturally, and that the system
reaches a quasi-equilibrium state in which the long-wavelength part of
the spectrum has a high effective temperature. This enhances baryon
number violation and makes baryogenesis more efficient.
}

\section{Introduction}
If CP is violated at high energies by some physics
beyond the standard model, the electroweak theory and the standard
Big Bang scenario seem to contain all the ingredients for explaining
the baryon asymmetry of the universe.\cite{Kuzmin:1985mm} 
The necessary out-of-equilibrium
conditions are provided by the dynamics of the phase transition and 
the last one of the three Sakharov conditions,\cite{Sakharov:1967dj} baryon
number violation, is satisfied by sphaleron processes, which change the
Chern-Simons number and consequently, due to a quantum anomaly, also
the baryon number.

However, although sphaleron processes become less frequent after the phase
transition, they don't disappear completely. Instead, their rate
is proportional to $\exp(-M_{\rm sph}/T)$, where 
$M_{\rm sph}\propto\phi$ and $\phi$ is the expectation value of
the Higgs field, and unless $\phi$ is large enough, 
the baryon asymmetry generated in the transition
can be washed out. This can only be avoided if the transition is
strongly enough first order so that the discontinuity of the Higgs field is
$\Delta\phi\gsim T$.

In the minimal standard model, the Higgs mass $m_H$ is the only unknown
parameter, and lattice simulations\cite{Kajantie:1997qd} 
have revealed that, whatever
its value, the transition is not strong enough. In more complicated
models, such as MSSM, there are more unknown parameters and this
freedom makes it possible to satisfy the constraint, but only barely.

In an alternative scenario
proposed recently by two 
groups,\cite{Krauss:1999ng,Garcia-Bellido:1999sv} 
the electroweak baryogenesis takes place during a period of
preheating after inflation.
This requires that inflation ends at an energy scale
that is below the electroweak scale and that 
a large fraction of the energy of the inflaton is transferred
rapidly to the standard model fields by a parametric 
resonance.\cite{kofman94} 
In the resulting non-equilibrium power spectrum, all the
fermionic fields
and the
short-wavelength modes of the bosons
are practically in vacuum, but the
long-wavelength bosonic modes have a high energy density.
The sphaleron rate depends strongly on the temperature of these
long-wavelength modes and is therefore very high, and the
out-of-equilibrium
processes can generate a large baryon asymmetry very quickly. Eventually the
system equilibrates and the effective temperature decreases by a rate
determined by the decay rate of gauge bosons into fermions,
$\Gamma\sim 1$~GeV. The final temperature $T_{\rm reheat}$ is
determined by the initial energy density and provided that it is low
enough, $T_{\rm reheat}\lsim 0.5 T_c$, 
the sphaleron rate becomes negligible and the baryon washout
is avoided.

In this talk, I will discuss the recent numerical 
simulations,\cite{Rajantie:2000fd} in which this 
process was studied in the Abelian
Higgs model.

\section{Simulations}
Instead of considering any particular model of inflation, 
we simply assume that the
inflaton interacts mostly with the Higgs, and that from the point of
view of the gauge fields, we can describe the Higgs and the inflaton 
by a single scalar
degree of freedom, which has a large energy density in its
long-wavelength
modes, which is realized by giving the scalar field $\phi$ a large
initial value $\phi_0$.
Since very little is known about the details of the inflaton, it is
difficult to improve this approximation.

When inflation ends, the inflaton field is typically still quite far
away from its minimum and contains a large amount of energy. Because
inflation dilutes all inhomogeneities, this energy is concentrated
in the very long-wavelength modes. Moreover, all the standard model
fields are in vacuum.

Instead of the full standard model gauge group, we used only a single
Abelian gauge field, because that makes the simulation much
simpler and is unlikely to change the qualitative behaviour.
The Lagrangian of our model is
\begin{equation}
L=
-\frac{1}{4}F^{\mu\nu}F_{\mu\nu}
+(D^\mu\phi)^*D_\mu\phi-\lambda(|\phi|^2-v^2)^2,
\label{ownlag}
\end{equation}
where the gauge covariant derivative is 
$D_\mu\phi=\partial_\mu\phi+ieA_\mu\phi$, and 
$F_{\mu\nu}=A_{\nu;\mu}-A_{\mu;\nu}$.

Since the occupation numbers of the long-wavelength modes will be
high, the dynamics of the system can be approximated by the classical
equations of motion. However, the quantum vacuum fluctuations are
important
as seeds for the parametric resonance, and therefore we approximate
them by adding Gaussian fluctuations with the same two-point function
as in the quantum vacuum, i.e.
\begin{equation}
\langle \phi^*(\vec{k})\phi(\vec{k}')\rangle
=
\frac{1}{2\omega(\vec{k})}(2\pi)^3\delta^{(3)}(\vec{k}-\vec{k}').
\end{equation}
In a sense, this means that the quantum effects are approximated to
leading order in perturbation theory.

In the full electroweak case, the effective temperature of the
long-wavelength modes decreases mostly because the Higgs and gauge
bosons decay into fermions. We approximate this by letting the universe
expand according to
$a=\sqrt{1+2Ht}$ 
at the rate $H=\dot{a}/a\approx 0.7$~GeV$\sim \Gamma$. This has the
effect
of reducing the energy in the long-wavelength modes, and if we use
the conformal time coordinate
$\eta$ defined by $d\eta\equiv dt/a$,
it appears simply as a
changing mass term for the Higgs field
\begin{equation}
m^2(t)=-2\lambda v^2a^2+\partial^2_\eta a/a.
\end{equation}

In the simulation, we used a $240^3$ lattice with lattice spacing
$\delta x/a=1.4$~TeV$^{-1}$ and time step $\delta t/a=0.14$~TeV$^{-1}$.
The couplings were $e=0.14$, $\lambda=0.04$ and $v=246$~GeV. The
initial value of the Higgs field was $\phi_0=1$~TeV.
\section{Results}
\begin{figure}[t]
\epsfxsize=21pc 
~~~~~\epsfbox{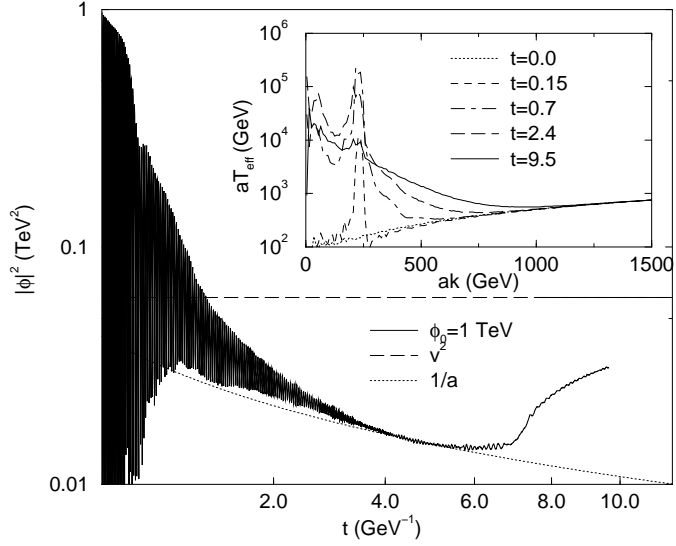} 

\caption{
The time evolution of $|\phi|^2$ on on a $240^3$ lattice with the
initial condition $\phi_0=1$~TeV. The symmetry is restored until
$t\approx 7$~GeV$^{-1}$. The effective temperature of different
Fourier modes is shown in the inset. Long-wavelength modes equilibrate
at a high temperature, but short wavelengths remain in vacuum.
\label{fig:kuva}
}
\end{figure}

The time evolution of 
$|\phi|^2$
is shown in Fig.~\ref{fig:kuva}. 
(We have subtracted the dominant ultraviolet divergence 
$\langle|\phi|^2\rangle_{\rm div}\approx0.226/\delta x^2$.) This model
does not have any local order parameter in the rigorous sense, and
$|\phi|^2$ in particular is non-zero in both the symmetric and the
broken phase. However, Fig.~\ref{fig:kuva} shows that
until $t\approx 7$~GeV$^{-1}$,
$|\phi|^2$ decreases roughly as $a^{-1}$,
indicated by the dotted line, which strongly suggests that no Higgs
condensate is present and that the system is in the symmetric phase.
After that,
the condensate develops, and $|\phi|^2$ starts to approach its
vacuum expectation value.

In the inset of Fig.~\ref{fig:kuva}, we have plotted the power
spectrum of the electric field in terms of the effective
temperature as a function of momentum at various times during the time
evolution. The definition of $T_{\rm eff}$ is
\begin{equation}
T_{\rm eff}(k)=\frac{1}{2}|E^{\rm T}_{i}(k)
|^2\frac{d^3k}{(2\pi)^3},
\end{equation}
where the superscript T indicates the transverse component
of the electric field;
the longitudinal component is fixed by the Gauss law.
In thermal equilibrium at
temperature $T$, $T_{\rm eff}(k)=T$ for every $k$. 
We can see that initially, the
power spectrum develops a sharp peak, which later spreads, and the
power spectrum reaches a quasi-equilibrium form in which the
long-wavelength modes $k\lsim k_*\sim e\phi_0$ are 
at a high effective temperature $T_{\rm eff}\sim\phi_0/e$.
The short-wavelength modes with $k\gsim k_*$ remain
in vacuum, where the effective temperature due to vacuum quantum
fluctuations
is $T_{\rm eff}=k_*/2$.
Because of the expansion of the space, the cutoff scale and the
effective momentum decrease as $a^{-1}$, but apart from that,
the form of the
power spectrum remains qualitatively the same until the end of our
simulations.

The form of this quasi-equilibrium power spectrum is crucial for the
scenario of electroweak baryogenesis at preheating. The sphaleron rate
is proportional to $T^4$ and is only sensitive to the long-wavelength
modes, and as their temperature is high, the baryon number violation
is very strong. Furthermore, $\langle\phi^2\rangle$ and $\langle
A_i^2\rangle$ also have much larger values than they would have in
equilibrium with the same energy density and
since the effective mass of the Higgs
field is given by
\begin{equation}
m_\phi^2\approx-2\lambda v^2+4\lambda\langle\phi^2\rangle
+e^2\langle A_i^2\rangle,
\end{equation}
the symmetry restoration is much easier.\cite{Rajantie:2000fd,kofman96} 
Thus it may be possible to
generate the baryon asymmetry during a period of this ``non-thermal''
symmetry restoration, and since the effective 
temperature decreases much faster
than
the baryons can decay, the baryon asymmetry will quickly freeze in.

Our results show that the qualitative behaviour of gauge-Higgs
models is compatible with electroweak baryogenesis at preheating.
In order to test the scenario quantitatively, we are currently
working on simulations in the SU(2)$\times$U(1) theory.\cite{future}

\section*{Acknowledgments}
I would like to thank E.J.~Copeland and P.M.~Saffin for collaboration on
this subject, and PPARC and the University of Helsinki for financial
support.  This work was conducted on the SGI Origin platform using COSMOS
Consortium facilities, funded by HEFCE, PPARC and SGI.

\end{document}